\documentclass[useAMS,usenatbib]{mn2e}
\usepackage{graphicx}
\begin{document}

\title[Identification of a 12--17 day timescale in X-ray observations of GRS 1915+105]{Identification of a 12--17 day timescale in X-ray observations of GRS 1915+105}
\author[J.~Greenhough, S.~C.~Chapman, S.~Chaty, R.~O.~Dendy and G.~Rowlands]{J.~Greenhough$^{1}$\thanks{Email: greenh@astro.warwick.ac.uk}, S.~C.~Chapman$^{1}$\thanks{Email: sandrac@astro.warwick.ac.uk}, S.~Chaty$^{2}$\thanks{Email: s.chaty@open.ac.uk. Present address: Service d'Astrophysique, DSM/DAPNIA/SAp, CEA/Saclay, Bât. 709
L'Orme des Merisiers, F-91 191 Gif-sur-Yvette, Cedex, France (chaty@discovery.saclay.cea.fr)}, R.~O.~Dendy$^{3,1}$\thanks{Email: richard.dendy@ukaea.org.uk} and G.~Rowlands$^{1}$\thanks{Email: g.rowlands@warwick.ac.uk}.\\ 
$^{1}$Department of Physics, University of Warwick, Coventry CV4 7AL, UK\\
$^{2}$Department of Physics and Astronomy, The Open University, Walton Hall, Milton Keynes MK7 6AA, UK\\
$^{3}$Euratom/UKAEA Fusion Association, Culham Science Centre, Abingdon, Oxfordshire, OX14 3DB, UK}

\date{Accepted . Received .}
\pagerange{\pageref{firstpage}--\pageref{lastpage}} \pubyear{2002}

\maketitle

\label{firstpage}

\begin{abstract}
Measurement of the energy emitted from accreting astrophysical systems provides an observational constraint on the plasma processes that may be operating within the disk. Here we examine the continual time variation over the past six years of the total X-ray flux from the microquasar GRS 1915+105. The application of differencing and rescaling techniques to RXTE/ASM data shows that the small amplitude fluctuations scale up to 12--17 days. A 17-day timescale in the X-ray fluctuations corresponds to half the measured binary orbital period of this system ($33.5\pm1.5$ days). While this may be coincidental, it is possible that these two timescales may be linked by, for example, a turbulent cascade in the accretion disk driven by a tidally-induced two-armed spiral shock corotating with the binary system.  Temporal scaling is found only in the ever-present small fluctuations, and not in the intermittent larger-amplitude fluctuations. This is consistent with the basic model for this source which consists of a steady, cold outer disk and an unstable inner disk. 
\end{abstract} 


\section{Introduction}\label{intro}
GRS 1915+105 was discovered in 1992 by the WATCH All-sky X-ray Monitor on board the GRANAT satellite \citep{castro1}, and lies in the Galactic plane at a distance of 11--12 kpc \citep{mirabel1,fender}. The extreme variability of the X-ray flux from its accretion disk, first noticed by \citet{castro2}, coupled with its jet emission \citep{mirabel1,mirabel2,dhawan,chaty}, makes it an ideal source for the study of accretion physics. At $14\pm4$ solar masses \citep[hereafter referred to as G2001]{greiner}, it can thus be considered as a Galactic stellar-mass analogue of a massive extragalactic black hole system such as a quasar, hence the term microquasar. The importance of microquasars as accessible astrophysical laboratories is discussed by \citet{mirabel2}; for example, the proportionality of dynamic timescales to mass implies that thousand-year events in a quasar could be studied over minutes in a microquasar (see for example \citet{mirabel3}).  
  
Spectral states, quasi-periodic oscillations (QPOs), and the rich temporal variability of GRS 1915+105 are the focus of much research (see for example \citet{nay,bell,klein,rod,ueda}; and references therein). Before the measurements by G2001, mass estimates for GRS 1915+105 made use of the temporal and spectral similarity to microquasar GRO J1655-40 \citep{grove}, another Galactic X-ray source with superluminal radio jets \citep{zang} together with a binary mass function appropriate for a black hole \citep{bailyn}. The radial velocity measurements performed by G2001 determined a mass of $14\pm4$ solar masses, leading the authors to suggest that the X-ray variability can be accounted for by instabilities in a radiation-pressure-dominated accretion disk near the Eddington limit. The observations of G2001 also constrain the mass of the donor star to $1.2\pm0.2$ solar masses, and its orbital period around the black hole to $33.5\pm1.5$ days.

There is now much evidence for the basic model of a cold, steady outer disk and an unstable, radiation-pressure-dominated inner disk in GRS 1915+105 (see for example \citet{bellonia,bellonib} and references therein). Hence there are two emitting regions to consider: an inner region that empties and refills on timescales of seconds, perhaps due to a rapid local viscous-thermal instability \citep{lightman,bellonia,janiuk}; and a steadier outer region varying only over the much longer timescale $t_{visc}$ associated with viscous dissipation of bulk plasma flows. We may therefore expect $t_{visc}$ to emerge as the longest timescale over which correlation may be present in observations of the total luminosity; a recent study suggests $t_{visc}\sim10$ years \citep{vilhu}. 

In this paper we apply the technique of differencing and rescaling to the GRS 1915+105 X-ray data, and show that this implies the existence of a fundamental timescale for the system in the range 12--17 days. The technique of differencing and rescaling was used by \citet{she} and \citet{shezag} to study a physical model of intermittency in turbulence, and by \citet{mantegna} to investigate fluctuations in the value of a financial index. More recently, the method has been used by \citet{greenh} to compare accreting with non-accreting X-ray sources, and by \citet{hnat} to quantify turbulence in the solar wind. In summary, we first construct a set of differenced series $Z\left(t,\tau\right)$ from the original series $y(t)$, for a range of values of the time-lag $\tau$:
\begin{equation}
Z\left(t,\tau\right)=y\left(t\right)-y\left(t-\tau\right).\label{differ}
\end{equation}
A set of PDFs is then calculated for the amplitude of the differenced series $Z$, one for each value of $\tau$, denoted by $P\left(Z,\tau\right)$. The best-fit slope of $\log P\left(0,\tau\right)$ against $\log\tau$ defines the scaling exponent $m$. We then seek a common functional form of these PDFs by rescaling both axes such that 
\begin{equation}
Z\rightarrow Z\tau ^{m}=Z_{s}\hspace{5mm}\mathrm{and}\hspace{5mm}P\rightarrow P\tau ^{-m}=P_{s}.\label{rescale}
\end{equation}
If the separate PDFs for values of $\tau$ up to some $\tau_{max}$ collapse onto one curve, this implies that the X-ray fluctuations are controlled by a single physical process on timescales below $\tau_{max}$.  

This technique enables us to test in particular for the joint presence of temporal scaling and non-Gaussianity \citep{hastings,kantz,sornette}, which together are strong indications of correlated processes such as turbulence \citep{bohr}. GRS 1915+105 is ideal for this type of statistical analysis in that (i)~it remains active on very long time scales, and (ii)~this time scale is very well separated from those associated with the inner disk.  

\section{Scaling of GRS 1915+105 X-ray data}
\subsection{Summary of the data}
\begin{figure}
	\includegraphics[width=84mm]{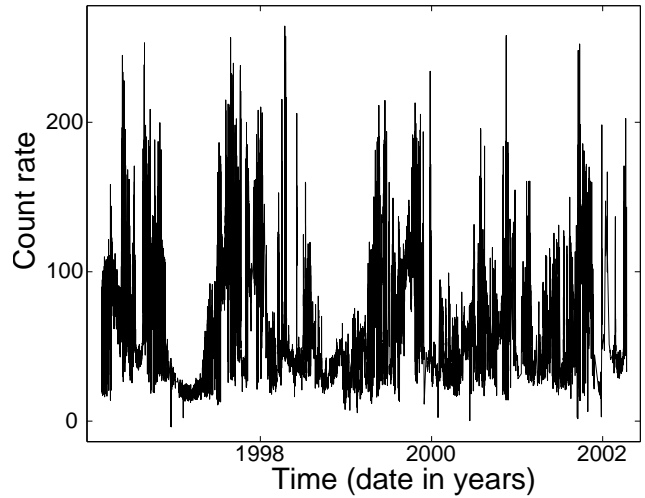}
	\caption{X-ray time-series for GRS 1915+105, 1996 Feb.~20 -- 2002 Apr.~12 (MJD 50133--52376); mean sampling interval 94 minutes. There are 32,800 data points, each of which is the number of counts in the range 1.3--12.2 keV observed in a 90-second interval. Data obtained by ASM on RXTE \citep{rxte,asm}.}\label{data}
\end{figure}

The raw X-ray data, provided by the All-Sky Monitor (ASM) on board the RXTE satellite \citep{rxte}, are held at the Goddard Space Flight Center (GSFC) and can be accessed via their website\footnote[1]{http://heasarc.gsfc.nasa.gov/docs/xte/asm\_products.html}. Calibration is undertaken by the ASM/RXTE team and the processed data are freely accessible on their website\footnote[2]{http://xte.mit.edu/XTE/asmlc/ASM.html} \citep{asm}, where there are 32,800 observations over the period 1996 February 20 to 2002 April 12, which we plot in Fig.~\ref{data}. 
Each point represents the total X-ray flux (measured by the number of counts during periods that last 90 seconds) integrated over the range 1.3--12.2 keV. The breakdown into three energy bands (1.3--3, 3--5 and 5--12.2 keV) is also available; we have carried out separate analyses of these components, which yield very similar results to those for the total flux which we present here. The source luminosity is sufficiently high to permit the neglect of instrument thresholds, uncertainties, and other sources in the field of view. Sampling intervals between the 90-second X-ray counting periods are distributed with a mean of 94 minutes, and 90\% of the intervals are below 193 minutes; the exact observation times are known and therefore we do not assume even sampling in our analysis. When differencing the X-ray intensities to construct $P(Z,\tau)$, we exploit the fact that each measured intensity is obtained at a precise time that is stated in the dataset to within 0.1s. We use these sampling times to find intensities that were observed within $2^{\pm 0.1}\tau$ of the chosen value of $\tau$. While the ASM data is not filtered prior to sampling, and hence the true frequency spectrum is distorted by aliasing, the differencing technique uses only time domain information that accurately represents the true signal at the given times. We do not attempt any kind of interpolation.
 
\subsection{Results}
\begin{figure}
	\includegraphics[width=84mm]{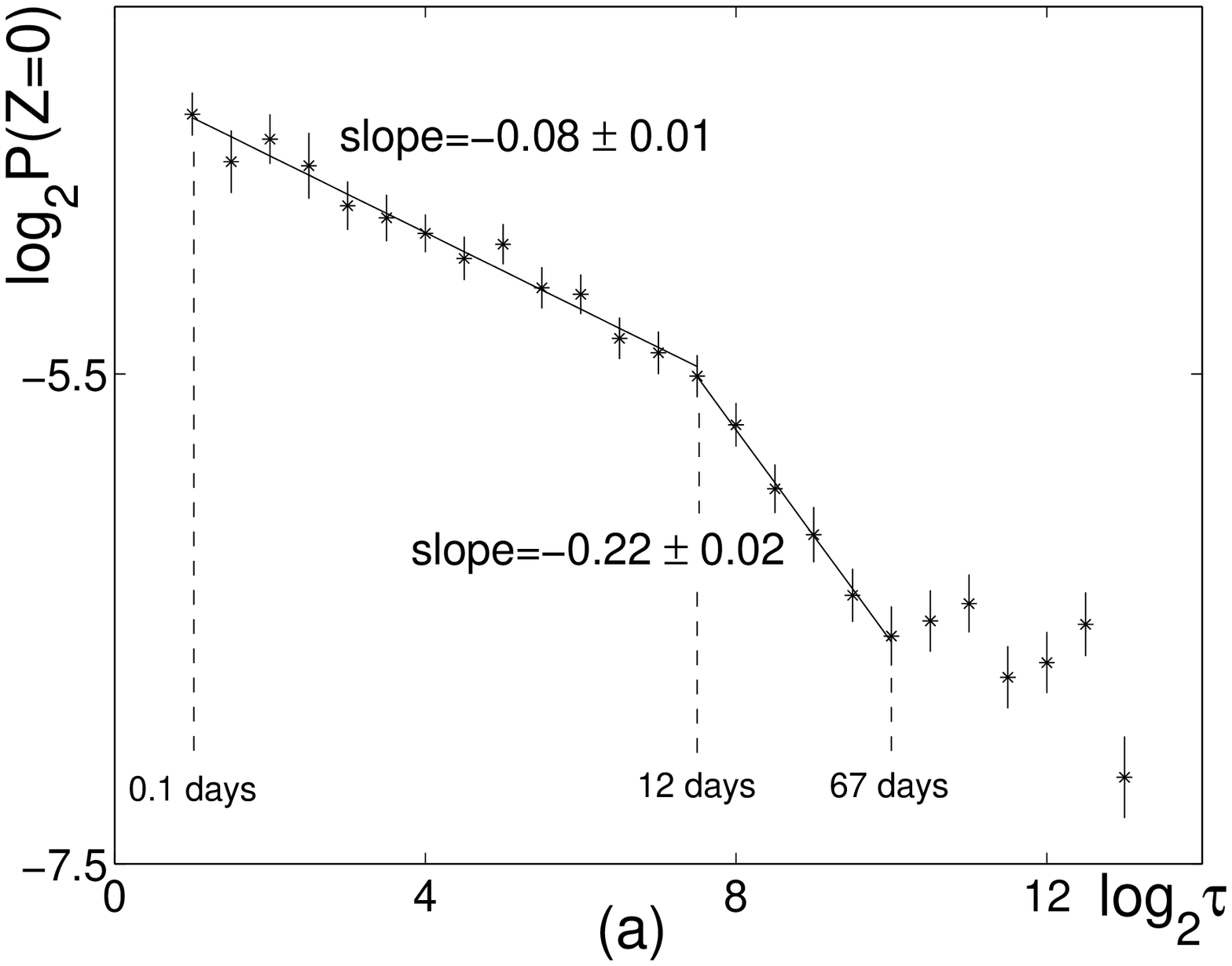}
	\includegraphics[width=84mm]{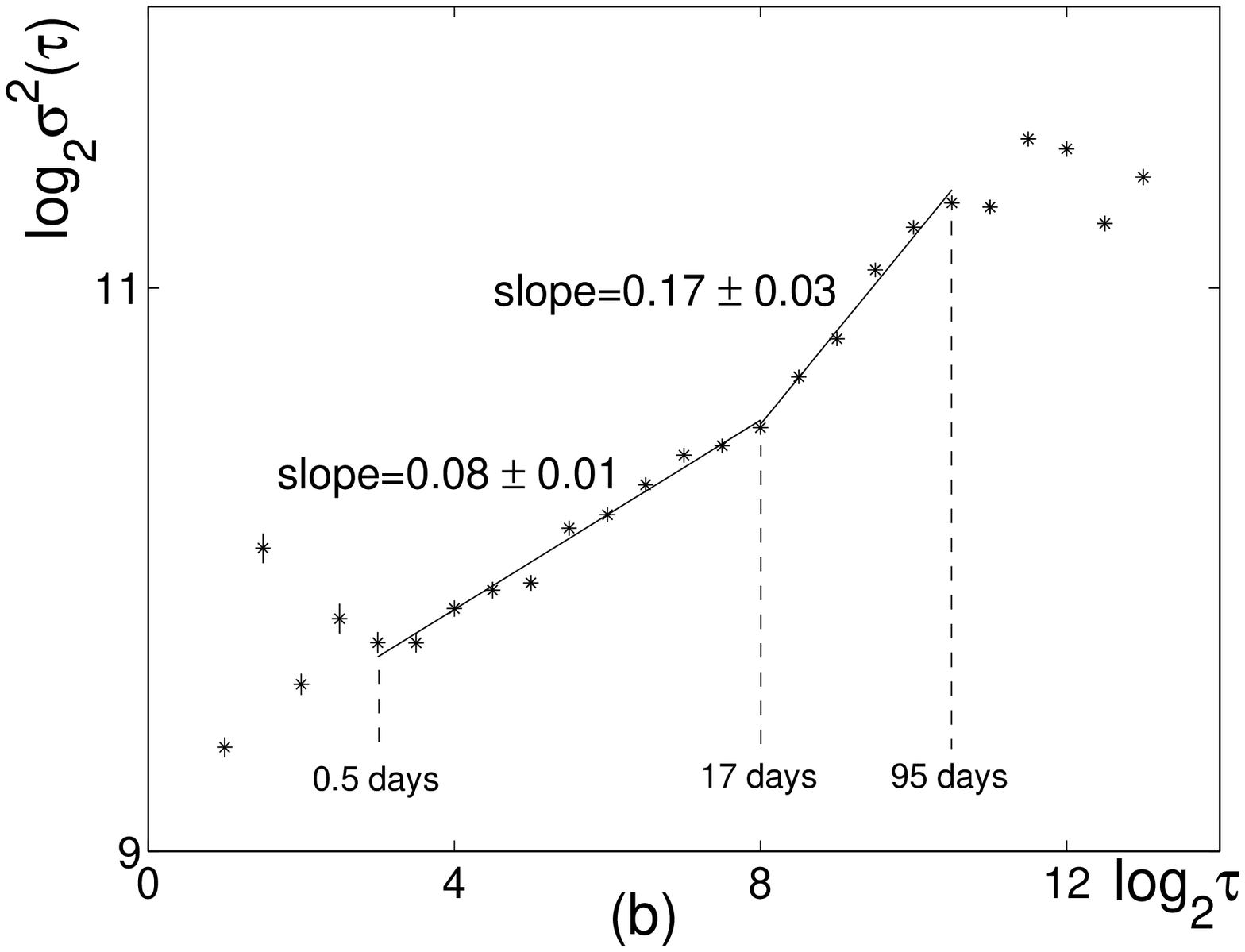}
	\caption{(a)~Dependence of the probability of the smallest-amplitude fluctuations (differences), $P(0,\tau)$, on differencing time interval $\tau$. This is a logarithmic plot of $P(0,\tau)$ against $\tau=j\tau_{0}$, where $j$ steps up in integer powers of two and $\tau_{0}=94$ minutes. Vertical error bars indicate a factor of $\pm 1/\sqrt{N}$, where $N=$ number of pairs of data points separated by $\tau$ that are used to calculate $P(0,\tau)$; horizontal error bars of $\pm2^{0.1}$ indicate the range of $\tau$ separating data points used to obtain $P(Z,\tau)$. The slope $m=-0.08$ is used to derive Fig.~3. (b)~Dependence of the variance, $\sigma^{2}(\tau)$, of the set of differences $(Z,\tau)$ on differencing time interval $\tau$; error bars are calculated as in (a). The slopes of the solid lines in both plots are calculated from linear regression with 95\% confidence intervals. Taken together, the plots suggest that scaling extends from 0.5 days to 12--17 days.}\label{p0}	
\end{figure}

In Fig.~\ref{p0}(a) we plot the probability of the smallest fluctuations, $P(0,\tau)$, as a function of $\tau$.  The straight lines on the double logarithmic axes indicate that $P(0,\tau)$ is a power-law function of $\tau$ up to $\tau\approx2^{7.5}$. This corresponds to a timescale of 12 days, since this value of $\tau$ represents $2^{7.5}$ times the mean sampling interval of 94 minutes. This break in scaling appears to be a robust feature of the data. For example, Fig.~\ref{p0}(b) shows the variance $\sigma^{2}(\tau)$ of the set of all fluctuations for a given value of $\tau$; that is, the variance of the differenced data from which the PDFs $P(Z,\tau)$ are calculated. It implies the existence of scaling for $\tau$ up to $2^{8}$, corresponding to 17 days. Our key result is that there is a clear break in scaling between the 7.5th and 8th octaves, corresponding to a timescale of 12--17 days. We emphasise that 12--17 days is not a periodicity in the X-ray luminosity but a timescale at which a break in the scaling of its statistical properties is observed. In Figs.~\ref{p0}(a,b) a second scaling region emerges between octaves 7.5 and 10.5. This represents a rather restricted range of timescales, however, and will not be considered further in this paper.  

\begin{figure}
	\includegraphics[width=84mm]{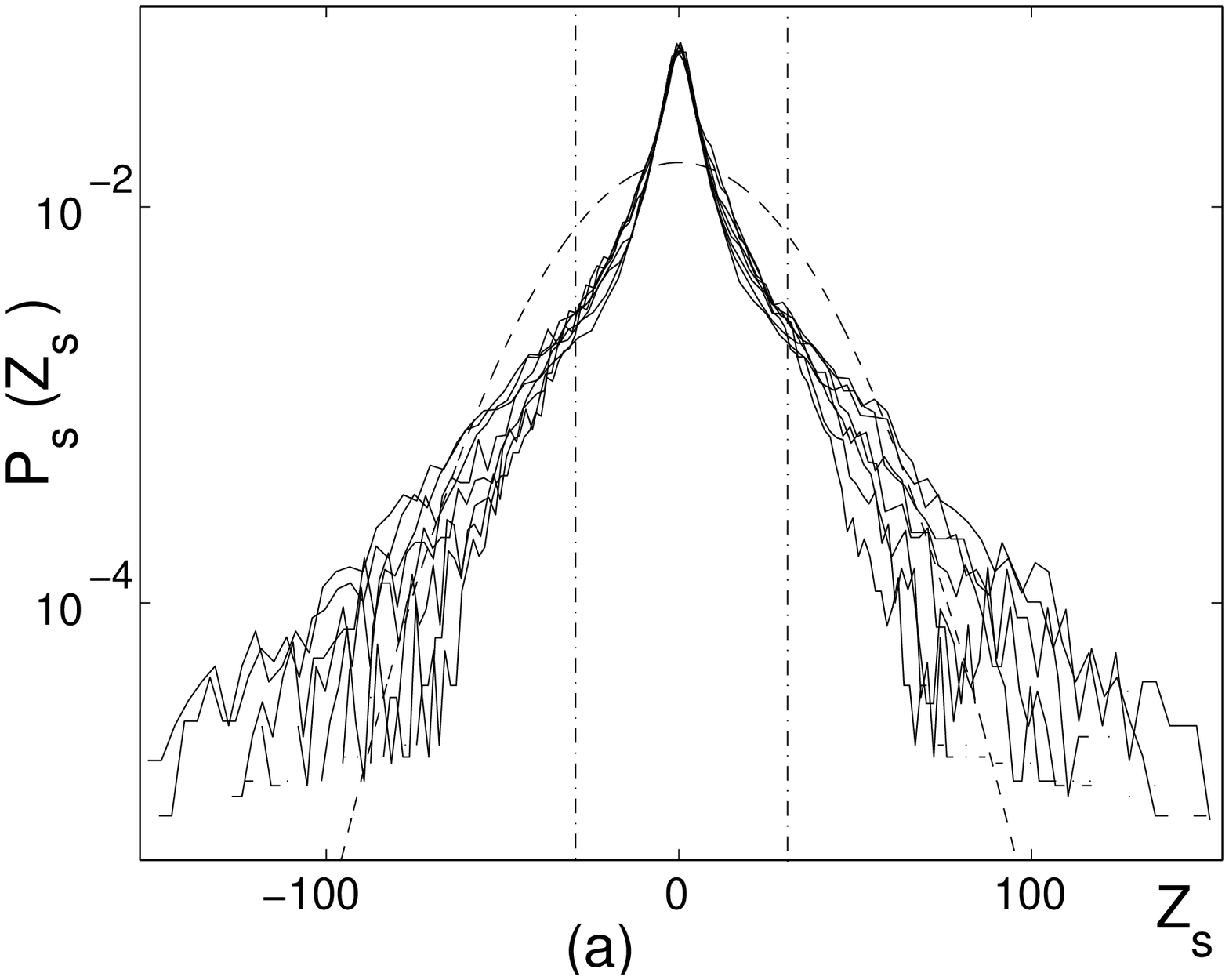}
	\includegraphics[width=84mm]{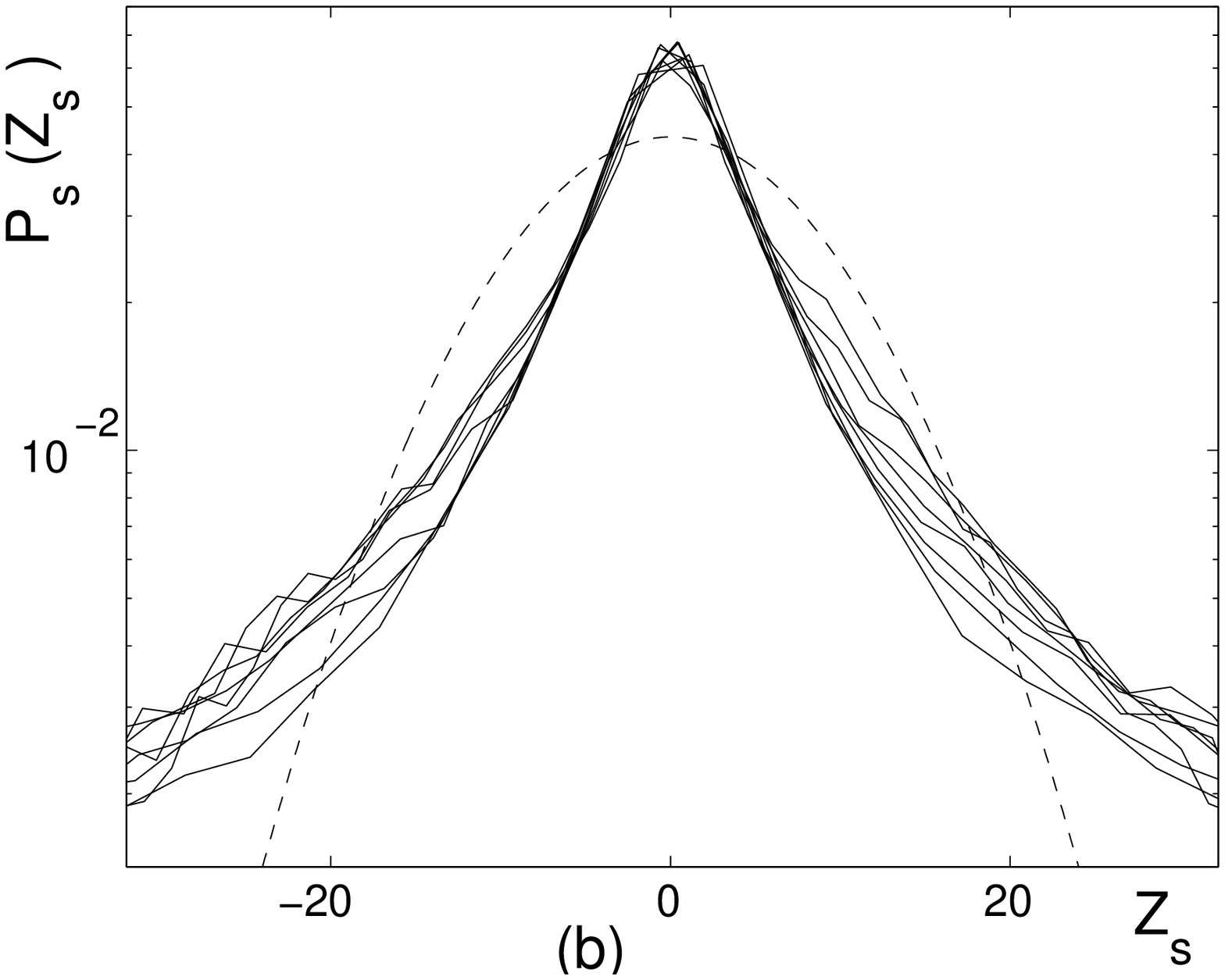}
	\caption{(a)~Rescaled PDFs of differenced time-series, where $Z_{s}$ denotes the differences rescaled according to $\tau$ and $m$ as at Eq.~\ref{rescale}. Curves are plotted for time intervals $\tau=j\tau_{0}$ where $j$ steps up in integer powers of two and $\tau_{0}=94$ minutes; $j_{max}=8$ ($=17$ days). The PDFs collapse onto one non-Gaussian curve between the dot-dash lines at $\mid Z_{s}\mid=30$, indicating that only the small fluctuations exhibit temporal scaling. For comparison purposes, the dashed line is the Gaussian fit to curve $j=1$ for all $Z_{s}$. (b)~Peak region of (a) enlarged to display detail, with Gaussian (dashed line) fitted to curve $j=1$ for $\mid Z_{s}\mid <30$ only.}\label{8scale}
\end{figure}

We next use the slope $m=-0.08$ of the linear fit over the lower range of timescales in Fig.~\ref{p0}(a) to carry out rescaling, as described at Eq.~\ref{rescale}, for fluctuations of size $Z$. Figure~\ref{8scale}(a) shows a collapse onto one curve for $\mid Z_{s}\mid < 30$, implying that small fluctuations follow the same scaling, but large fluctuations beyond $\mid Z_{s}\mid\approx30$ clearly do not scale in this way. In Fig.~\ref{8scale}(b) we show that the scaling r\'{e}gime indicated in Fig.~\ref{8scale}(a) is peaked, long-tailed, and definitely non-Gaussian, which is suggestive of turbulent behaviour \citep{bohr}.

\begin{figure}
	\includegraphics[width=84mm]{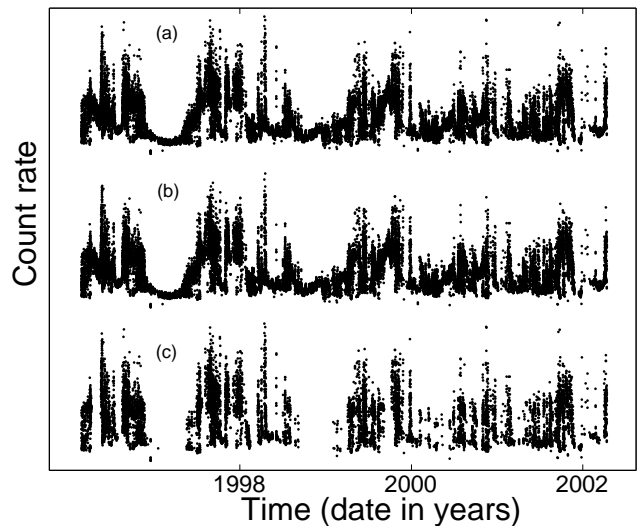}
	\caption{X-ray time-series plotted using dots to identify the origin of  fluctuations of different sizes; (a)~all fluctuations (all data points), (b)~small fluctuations only (between the dot-dash lines in Fig.~\ref{8scale}), (c)~large fluctuations only (outside the dot-dash lines in Fig.~\ref{8scale}). This plot shows that only the small fluctuations originate from all parts of the time-series, and Fig.~\ref{8scale} shows that only these fluctuations exhibit scaling.}\label{flucs}
\end{figure}

We now show that the small fluctuations that exhibit scaling are continually present, whereas the large fluctuations are not. Figure~\ref{flucs} provides representations of: (a)~the full X-ray time-series for GRS 1915+105; (b)~the small fluctuations that lie between the dot-dash lines ($\mid Z_{s}\mid<30$) in Fig.~\ref{8scale}; and (c)~the large fluctuations ($\mid Z_{s}\mid>30$) only. Dots are the data points at which the fluctuations terminate, that is the latter points of the pairs that are differenced to construct $P(Z,\tau)$. Thus, for example, in Fig.~\ref{flucs}(a) there is a dot for every raw data point, whereas in (b) we plot one dot only for each pair of intensities whose rescaled difference $Z_{s}$ lies between $Z_{s}=\pm 30$. It is clear that the small scaling fluctuations arise from all parts of the time-series, whereas the larger non-scaling fluctuations are sometimes absent (notably for long periods in 1997 and 1998--99). 

\section{Discussion and Conclusions}\label{discuss}
Using the differencing and rescaling technique, 12--17 days emerges as a resilient timescale in the total X-ray luminosity (1.5--12 keV) of GRS 1915+105, and is shown to originate in only the small luminosity fluctuations. This is consistent with the accepted model of a cold (low emission and variability) outer disk that is stable over (viscous) timescales well above those of the inner disk; only in the outer disk could processes evolve without interruption over periods of days.    
     
It is also intriguing that 17 days is approximately half the binary period ($33.5\pm1.5$ days) of this source. Although this may be coincidental, it is interesting to speculate on its possible physical significance. For example, this result lends some support to an interpretation in terms of tidally-driven waves which we now describe. \citet{lubow} review fluid experiments beginning in the 1970s, in which it was found that a dynamical instability can arise in a fluid whose base state is non-circular; \citet{goodman} originally applied this idea to tidally-distorted accretion disks in close binary systems. The instability is caused by a perturbation in the frame rotating with the binary at angular frequency $\Omega_{B}$. Trailing waves can couple via the tidal mode to generate a leading wave, which can then reinforce the original disturbance via the tidal distortion. \citet{vishniac} use more detailed disk models to confirm Goodman's findings, and \citet{ryu} show how the instability can become global (for other types of instability see, for example, \citet{tagger}). 

In the mode with the largest amplitude, the surface density varies slowly with radius and has an angle and time dependence given by $\sin(2\theta-2\Omega_{B}t)$ \citep{lubow}. Differential rotation within the disk shears this disturbance into a two-armed spiral shock that co-rotates with the binary; amplification caused by the inwardly-increasing velocity is balanced by dissipation \citep{blondin}. This tidally-induced shock wave has been shown numerically to be a robust feature of the accretion disks of binary systems \citep[and references therein]{blondin}, and their possible contribution to the light curves of intermediate polars was demonstrated by \citet{murray}. There exist, in addition, many observations of spiral shocks in binaries (see for example \citet{baptista} and references therein) and in AGN (\citet{rouan} and references therein, for example). It is plausible that the periodic stirring effect of the rotating shock could drive a turbulent cascade \citep{goodman,bohr} in which plasma is heated by viscous dissipation within vortices covering a range of spatial and temporal scales smaller than those of the driving process, and with non-Gaussian fluctuations. It is this dissipation, spatially integrated across the accretion disk, that drives the X-ray emission. Our observations are consistent with the hypothesis that in the present case of turbulence in a bounded system, the temporal scaling properties -- and in particular the upper timescale of 12--17 days -- persist in the X-ray signal. With two spiral arms each rotating with a period of $T_{B}=2\pi/\Omega_{B}$, one would therefore expect to observe temporal scaling up to $T_{B}/2$ in observations of the total luminosity. For GRS 1915+105, $T_{B}/2=(33.5\pm1.5)/2=17\pm1$ days.

Our principal result thus provides quantitative constraints on models for transport and heating in the accretion disk of GRS 1915+105, and our analysis also suggests possible physical mechanisms that may be at work. 

\section*{Acknowledgements}
We are grateful to Michel Tagger, James Murray, Nick Watkins (for drawing our attention to Blondin (2000)), David Tsiklauri and Bogdan Hnat for helpful suggestions. J.~G. acknowledges a CASE Research Studentship from the UK Particle Physics and Astronomy Research Council in association with UKAEA. This work was also supported in part by the UK DTI. S.~C. gratefully acknowledges support from grant F/00--180/A from the Leverhulme Trust, and G.~R. acknowledges a Leverhulme Emeritus Fellowship. Data provided by the ASM/RXTE teams at MIT and at the RXTE SOF and GOF at NASA's GSFC.

\label{lastpage}


\begin{thebibliography}{99}
\bibitem[\protect\citeauthoryear{Bailyn et al.}{1995}]{bailyn} Bailyn C.~D., Orosz J.~A., McClintock J.~E., and Remillard R.~A. 1995, Nature, 378, 157
\bibitem[\protect\citeauthoryear{Baptista et al.}{2002}]{baptista} Baptista R., Haswell C.~A., Thomas, G. 2002, MNRAS, 334, 198
\bibitem[\protect\citeauthoryear{Belloni et al.}{1997a}]{bellonia} Belloni T., M\'{e}ndez M., King A.~R., van der Klis M., and van Paradijs J. 1997, ApJ, 479, L145
\bibitem[\protect\citeauthoryear{Belloni et al.}{1997b}]{bellonib} Belloni T., M\'{e}ndez M., King A.~R., van der Klis M., and van Paradijs J. 1997, ApJ, 488, L109
\bibitem[\protect\citeauthoryear{Belloni et al.}{2001}]{bell} Belloni T.,  M\'{e}ndez M., and S\'{a}nchez-Fern\'{a}ndez C. 2001, A\&A, 372, 551
\bibitem[\protect\citeauthoryear{Blondin}{2000}]{blondin} Blondin J.~M. 2000, New Astronomy, Vol.~5, No.~1, 53
\bibitem[\protect\citeauthoryear{Bohr et al.}{1998}]{bohr} Bohr T., Jensen M.~H., Paladin G., and Vulpiani A. 1998, Dynamical Systems Approach to Turbulence (Cambridge, UK: Cambridge University Press)
\bibitem[\protect\citeauthoryear{Bradt et al.}{2001}]{asm} Bradt H.~V., Chakrabarty D., Cui W., et al. 2001, ASM Light Curves Overview (MIT, MA: ASM/RXTE team)
\bibitem[\protect\citeauthoryear{Castro-Tirado et al.}{1992}]{castro1} Castro-Tirado A.~J., Brandt S., and Lund S. 1992, IAU Circ. 5590
\bibitem[\protect\citeauthoryear{Castro-Tirado et al.}{1994}]{castro2} Castro-Tirado A.~J., Brandt S., Lund S. et al. 1994, ApJS, 92, 469
\bibitem[\protect\citeauthoryear{Caunt and Tagger}{2001}]{tagger} Caunt S.~E., and Tagger M. 2001, A\&A, 367, 1095
\bibitem[\protect\citeauthoryear{Chaty et al.}{2001}]{chaty} Chaty S., Rodr\'{\i}guez L.~F., Mirabel I.~F. et al. 2001, A\&A, 366, 1035
\bibitem[\protect\citeauthoryear{Dhawan et al.}{2000}]{dhawan} Dhawan V., Mirabel I.~F., and Rodr\'{\i}guez L.~F. 2000, ApJ, 543, 373
\bibitem[\protect\citeauthoryear{Fender et al.}{1999}]{fender} Fender R.~P., Garrington S.~T., McKay D.~J., et al. 1999, Mon.~Not.~R.~Astron.~Soc., 304, 865
\bibitem[\protect\citeauthoryear{Frank et al.}{1992}]{frank} Frank J., King A.~R., and Raine D.~J. 1992, Accretion Power in Astrophysics, 2nd Ed., (Cambridge, UK: Cambridge University Press) 
\bibitem[\protect\citeauthoryear{Goodman}{1993}]{goodman} Goodman J. 1993, ApJ, 406, 596
\bibitem[\protect\citeauthoryear{Greenhough et al.}{2002}]{greenh} Greenhough J., Chapman S.~C., Chaty S., Dendy R.~O., and Rowlands G. 2002, A\&A, 385, 693 
\bibitem[\protect\citeauthoryear{Greiner et al.}{2001}]{greiner} Greiner J., Cuby J.~G., and McCaughrean M.~J. 2001, Nature, 414, L522
\bibitem[\protect\citeauthoryear{Grove et al.}{1998}]{grove} Grove J.~E., Johnson W.~N., Kroeger R.~A. et al. 1998, ApJ, 500, 899
\bibitem[\protect\citeauthoryear{Hastings and Sugihara}{1993}]{hastings} Hastings H.~M., and Sugihara G. 1993, Fractals: a User's Guide for the Natural Sciences (New York: Oxford University Press)
\bibitem[\protect\citeauthoryear{Hnat et al.}{2002}]{hnat} Hnat B., Chapman S.~C., Rowlands G., Watkins N.~W., and Farrell W.~M. 2002, GRL, 29(10), 10.1029/2001GL014587
\bibitem[\protect\citeauthoryear{Janiuk et al.}{2000}]{janiuk} Janiuk A., Czerny B., and Siemiginowska A., 2000, ApJ, 542, L33
\bibitem[\protect\citeauthoryear{Kantz and Schreiber}{1997}]{kantz} Kantz H., and Schreiber T. 1997, Nonlinear Time Series Analysis (Cambridge, UK: Cambridge University Press)
\bibitem[\protect\citeauthoryear{Klein-Wolt et al.}{2002}]{klein} Klein-Wolt M., Fender R.~P., Pooley G.~G., Belloni T., Migliari S., Morgan E.~H., and van der Klis, M. 2002, MNRAS, 331, 745
\bibitem[\protect\citeauthoryear{Lightman and Eardley}{1974}]{lightman} Lightman A.~P., and Eardley D.~M. 1974, ApJ, 187, L1 
\bibitem[\protect\citeauthoryear{Lubow et al.}{1993}]{lubow} Lubow S.~H., Pringle J.~E., and Kerswell R.~R. 1993, ApJ, 419, 758
\bibitem[\protect\citeauthoryear{Malamud and Turcotte}{1999}]{malamud} Malamud B.~D., and Turcotte D.~L. 1999, Advances in Geophysics, Vol.~40 (San Diego, CA: Academic Press)
\bibitem[\protect\citeauthoryear{Mantegna and Stanley}{1995}]{mantegna} Mantegna R.~N., and Stanley H.~E. 1995, Nature, 376, 46
\bibitem[\protect\citeauthoryear{Mirabel and Rodr\'{\i}guez}{1994}]{mirabel1} Mirabel I.~F., and Rodr\'{\i}guez L.~F. 1994, Nature, 371, 46
\bibitem[\protect\citeauthoryear{Mirabel et al.}{1998}]{mirabel3} Mirabel I.~F., et al. 1998, A\&A, 330, L9
\bibitem[\protect\citeauthoryear{Mirabel and Rodr\'{\i}guez}{1999}]{mirabel2} Mirabel I.~F., and Rodr\'{\i}guez L.~F. 1999, Annu. Rev. Astron. Astrophys., 37, 409
\bibitem[\protect\citeauthoryear{Murray et al.}{1999}]{murray} Murray J.~R., Armitage P.~J., Ferrario L., and Wickramasinghe D.~T. 1999, MNRAS, 302, 189
\bibitem[\protect\citeauthoryear{Nayakshin et al.}{2000}]{nay} Nayakshin S., Rappaport S., and Melia F. 2000, ApJ, 535, 798
\bibitem[\protect\citeauthoryear{Pringle and Rees}{1972}]{pringle} Pringle J.~E., and Rees M.~J. 1972, A\&A, 21, 1
\bibitem[\protect\citeauthoryear{Rodr\'{\i}guez et al.}{2002}]{rod} Rodr\'{\i}guez L.~F., Durouchoux Ph., Mirabel I.~F., Ueda Y., Tagger M., and Yamaoka K. 2002, A\&A, 386, 271
\bibitem[\protect\citeauthoryear{Rouan et al.}{1998}]{rouan} Rouan D. et al. 1998, A\&A, 339, 687
\bibitem[\protect\citeauthoryear{Ryu et al.}{1996}]{ryu} Ryu D., Goodman J., and Vishniac E.~T. 1996, ApJ, 461, 805
\bibitem[\protect\citeauthoryear{Shakura and Sunyaev}{1973}]{shak} Shakura N.~I., and Sunyaev R.~A. 1973, A\&A, 24, 337
\bibitem[\protect\citeauthoryear{She}{1991}]{she} She Z.-S. 1991, PRL, 66, 5, 600
\bibitem[\protect\citeauthoryear{She and Orszag}{1991}]{shezag} She Z.-S., and Orszag S.~A. 1991, PRL, 66, 13, 1701
\bibitem[\protect\citeauthoryear{Sornette}{2000}]{sornette} Sornette D. 2000, Critical Phenomena in Natural Sciences (Berlin: Springer-Verlag)
\bibitem[\protect\citeauthoryear{Swank et al.}{2001}]{rxte} Swank J.~H., Smale H.~P., Boyd P.~T., et al. 2001, RXTE Guest Observer Facility (GSFC, MA: RXTE GOF)
\bibitem[\protect\citeauthoryear{Ueda et al.}{2002}]{ueda} Ueda Y. et al. 2002, ApJ, 571, 918
\bibitem[\protect\citeauthoryear{Vilhu}{2002}]{vilhu} Vilhu O. 2002, A\&A, 388, 936
\bibitem[\protect\citeauthoryear{Vishniac and Zhang}{1996}]{vishniac} Vishniac E.~T., and Zhang C. 1996, ApJ, 461, 307
\bibitem[\protect\citeauthoryear{Zhang et al.}{1994}]{zang} Zhang S.~N., Wilson C.~A., Harmon B.~A. et al. 1994, IAU Circ. 6046
\end{thebibliography}
\end{document}